# Use of Digital Technologies in Public Health Responses to Tackle Covid-19: the Bangladesh Perspective

Samrat Kumar Dey[*1], Khaleda Mehrin[2], Lubana Akter[3] and Mshura Akter[4]

[1]School of Science and Technology (SST),
Bangladesh Open University (BOU), Gazipur-1705, Bangladesh
[2]Bangladesh Bureau of Educational Information and Statistics (BANBEIS),
Ministry of Education, Bangladesh
[3,4] Department of Computer Science and Engineering (CSE),
Dhaka International University (DIU), Dhaka-1205, Bangladesh

*ABSTRACT*

*This paper aims to study the fight against COVID-19 in Bangladesh and digital intervention initiatives. To achieve the purpose of our research, we conducted a methodical review of online content. We have reviewed the first digital intervention that COVID-19 has been used to fight against worldwide. Then we reviewed the initiatives that have been taken in Bangladesh. Our paper has shown that while Bangladesh can take advantage of the digital intervention approach, it will require rigorous collaboration between government organizations and universities to get the most out of it. Public health can become increasingly digital in the future, and we are reviewing international alignment requirements. This exploration also focused on the strategies for controlling, evaluating, and using digital technology to strengthen epidemic management and future preparations for COVID-19.*

## 1. Introduction

COVID-19, also known as 2019 novel coronavirus (2019-nCOV), is a new strain of coronavirus that was first identified in Wuhan, China, in December 2019 and has drastically spread throughout the whole world. Today, 220 Countries and territories worldwide have confirmed 209,326,903 cases according to WHO (World Health Organization) of the COVID-19 and death full in 4393412 [1]. In Bangladesh, we have around 1461998 confirmed cases of COVID-19 and a death toll of 25282. In March 2019, we had 3 cases of COVID-19 victims and were confirmed by the Institute of Epidemiology, disease control and Research (IEDCR) for the first time [2]. After that, the Government of Bangladesh ensured strict nationwide lockdown.

From March 18, the Government of Bangladesh has structured real-time RT-PCR testing in 2019. But still, it is deficient. However, the authorities have taken multipurpose steps against the viruses, including diagnosis of the suspected cases, quarantine of doubted people and isolation of infected patients, and local or regional lockdown—closure of all Government and enforcing social distaining ICU and Oxygen facilities. But the government have unravelled to make proper strategies, including contact tracing introducing antigen-based rapid detection kit and tailed to make a multi-disciplinary team to combat this disease. Again, lack of testing facilities and insufficient treatment services are the main challenge for Bangladesh to tackle this situation successfully. In Bangladesh, mainly young professionals and working people are mostly affected with COVID-19. IEDCR reported that 68% of COVID-19 positive cases were 21 to 50 years of age. Over such years, people on the other side constituted 21% of the total infected people [3]. The children and young are also at life risk because of new types of COVID-19. However, in Bangladesh, mainly Dhaka, Narayanganj, Gazipur, and Chittagong have the most confirmed cases of COVID-19. Since March 8 2019, our educational sectors have been closed. Our government is trying to cope with this situation. But it is on the unpredictable challenge of such a developing country disease. The world is running to discovery for the progress of the digital environment, where the world depends on the vast network of the internet. In 2021 G.N Rebecca et al. worked on some internet-based platforms used by different countries around Asia [4]. That research aimed to review the role of big data and digital technologies in controlling COVID-19 in Asia. Those researchers implemented different Digital tools, including real-time epidemiological dashboards, interactive maps of case location, mobile apps for tracing patients' contacts, and geofencing to monitor quarantine willingness. The researchers were visualized that those Digital Technologies have naturalized and emphasized the traditional public health measures for preclusion of SARS- CoV-2 spread in Asia. Researchers constantly explore their research process to widen their research area. In this Covid time, they find their research in another level, 2020 researcher S. Whitelaw et al. in another work on the digital base technology in COVID-19 pandemic planning and response [5]. The researcher focused on the covid vaccine or therapy and how to digital technology helped general people in this Covid time. They highlighted how to coordinated government efforts across the globe had been focused on containment and mitigation, with varying degrees of success. They also highlight how successful countries had adopted this application for pandemic planning, surveillance, testing, contact tracing, quarantine, and health care. J. Budd et al. researched in 2020 based on Digital technologies in the public-health response to COVID-19 [6]. In this paper, scientists worked on a noble purpose, and they were made focused on the worldwide, including population surveillance, case identification, contact tracing, and evaluation of interventions based on mobility data and communicating with the public. They discover the width of digital innovations for the public-health response to COVID-19 worldwide and their brackets and interruption to their implementation, including legal, ethical, and privacy interruption, as well as organizational and workforce barriers. Q. Wang et al. [7] made an interest in integrating digital technologies and public health to fight COVID-19 pandemic. Researchers had been focused based on healthcare in this covid-19 pandemic. Research fellows have found their research results: geographic distribution, discipline distribution, collaboration network, and vital topics of digital healthcare before and after the COVID-19 pandemic. They were discovered that these challenges mainly come from four aspects: 1. data delays, 2. data fragmentation, 3. privacy security, and 4. data security vulnerabilities. Researchers provided the future application prospects of digital healthcare. Along with the Government, personal awareness and assistance of non-government organizations, private organizations, researchers, doctors, industrialists, and international organizations are firmly required to moderate this highly infected disease. Primary data were collected and analysed using a pre-created Google Survey form having a pre-set questionnaire on the social distancing status of different districts. Secondary data on the total and daily laboratory tests confirmed positive cases, and death cases were extracted from publicly available sources to make predictions. Estimated the primary reproduction number (R0) based on the SIR mathematical model and predicted the probable fate of this pandemic in Bangladesh. Quarantine situations in different regions of Bangladesh were evaluated and presented. This research also provided tentative forecasts until May 31 2020 and found that approximately the predicted curve followed the actual curve. Estimated R0 values (6.924) indicated that the infection rate would be greater than the recovery rate. Furthermore, by calibrating the parameters of the SIR model to fit the reported data, we assume the ultimate ending of the pandemic in Bangladesh by December 2022.

## 2. METHODOLOGY

This issue is still exotic, and all the academic literature is unavailable for the different applications and websites. We followed different online content based on mobile applications and websites to attain our research objectives. This study observed that the content approach is widely accepted in research. The online content was detached through systematic coding and interpretation. The exploration strings were used to find the guidable online content from global aspects was "digital interventions and coronavirus," "digital intervention and COVID19", "information technology and coronavirus," "information technology and COVID-19", "digital services and coronavirus," "digital services and COVID-19", "artificial intelligence and coronavirus", "artificial intelligence and COVID19", "mobile application and data science and robot and coronavirus," "mobile application and data science and robot and COVID-19", "electronic health services and coronavirus," and "electronic health services and COVID-19". The analogous set of keywords associated with 'and Bangladesh' were also used to explore the content focusing on behalf of Bangladesh, an example, "digital intervention and coronavirus and Bangladesh," "digital intervention and COVID-19 and Bangladesh", and similar that. The research output was created more than 1000 sites contents. Then we dispelled that content which was discussed by the same topics. Especially that news article focused on the same digital/ICT application, but that topic was published in multiple newspapers. These online contents were not focused on our research objectives, and the English and Bengali wrote that content. This exploration discovered various informatics applications that played an important role in this pandemic situation in the base of the Bangladesh platform. In this research, we explore all the applications and websites by owning. After exploring those, we find the prospective data on how people got benefits by using those mobile applications and websites. We explore the data based on Bangladesh that used those applications and websites. We figure out those data in Figure 1.

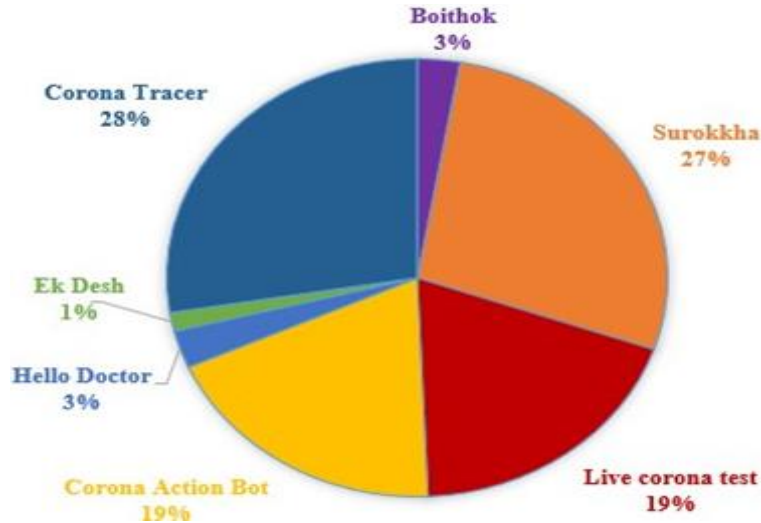

Figure 1. Users' perception towards different web and mobile based solution in Bangladesh

Then all the data were synthesized and analysed to explore the applications from a global perspective and in the context of Bangladesh. Finally, we conducted a comparative analysis between these two sets of exploring to assume the underlying facts of digital application in Bangladesh and future perspectives of accepting any new digital applications in Bangladesh to campaign the pandemic spread of COVID-19.

## 3. DIGITAL INTERVENTION TO TACKLE COVID-19

According to the report, on February 11 2021, around 47.61 million people are now depending on the internet, and internet usage is increasing day by day. With the increase of internet-user, mobile and social media users are also increasing. Around 165.8 million people use the mobile application, and 45.00 million people use a different social media site. In Bangladesh, this COVID pandemic situation this user of the internet, mobile, and different kind of application and the website number of users increased. Nowadays, for the betterment of the public and to lead a safe life, Bangladesh innovated different kinds of mobile applications, and websites are taking virtual place. Some digital innovations of mobile applications and websites for this covid pandemic following.

### 3.1. Boithok

This is a companion app for the Boithok platform, a web-based video conferencing platform [8]. The app is based on a zoom online meeting platform. The book is hosted in the National Data centre of Bangladesh Computer Council (BCC). And it's maintained by the BNDA team. This platform lets users share their necessities and feelings through perfect instant secure video conferencing that creates an accurate virtual meeting scenario. This platform is entirely secure, and users' data are entirely safe. This app is completely free to use but currently limited to government organizations. It is also available in the play store. No account is needed to join any conference. An account is needed to create a conference. The host can control the access to the conferences with a password—a feature of unlimited messaging and a private messaging option. No downloads are required to join the conversation. Users can directly join Boithok conferences within their browsers as well. A dedicated desktop version app is also available. Recording and sharing: A host can easily record their meeting, and any participant can share their screen with audio.

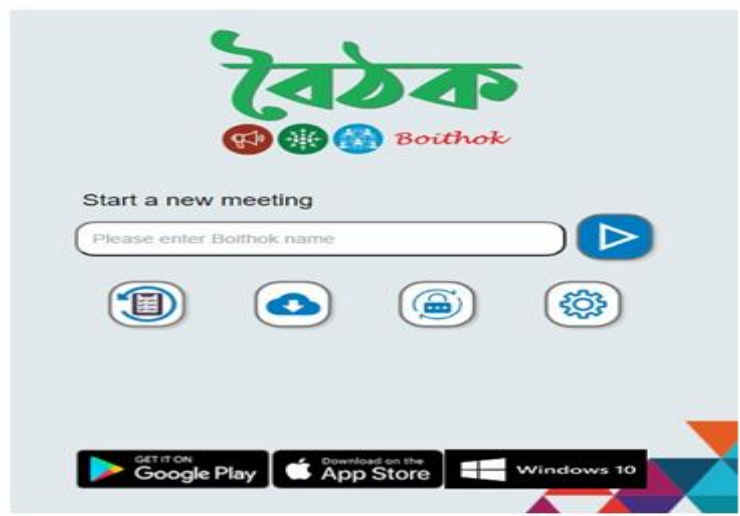

Figure 2. User Interface of Boithok application.

### 3.2. Surokkha

To distribute the COVID-19 vaccine among the people of Bangladesh, the ICT Division of Bangladesh has come up with a web portal and mobile application to proceed with the initial registration process [9]. Surokkha provides the facility to register for vaccination for the people of Bangladesh. If anyone wants to register for the COVID-19 vaccine, they must provide the National Identification number to verify. The following information is being captured from this application.

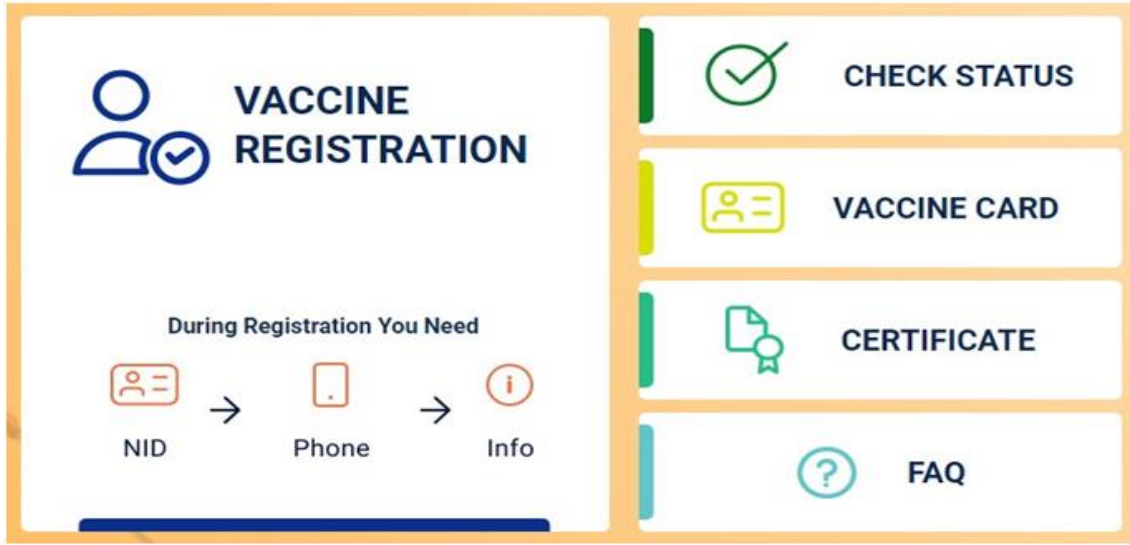

Figure 3. User Interface of Surokkha application.

- National Identification Number
- Date of Birth
- Mobile Phone number
- Comorbidity
- Desired Address for Vaccination Centre
- User consent for receiving the vaccine

The application verifies the user by sending OTP to the given mobile-phone number and registering. Registrants can check their application status, download the Vaccine Card and download the Certificate later. About last eight months by helped by Surokkha application about 31012179 people take was 1st doze and 13285707 people were taken 2nd doze of the COVID-19 vaccine in Bangladesh. Linear models are simple way to predict output using a linear function of input features.

## 3.3. Live COVID-19 Test

This is an extensive data analysis tool where the users will be able to take a 30-seconds quiz to answer questions that will determine the susceptibility of Coronavirus in Bangladesh. The objective of this tool is to have a bird' eye view of the status of the whole country, generate a considerable amount of data that was analysed to make strategic decisions. The Government will also use this to assess situations of communities, which will help take critical measures.

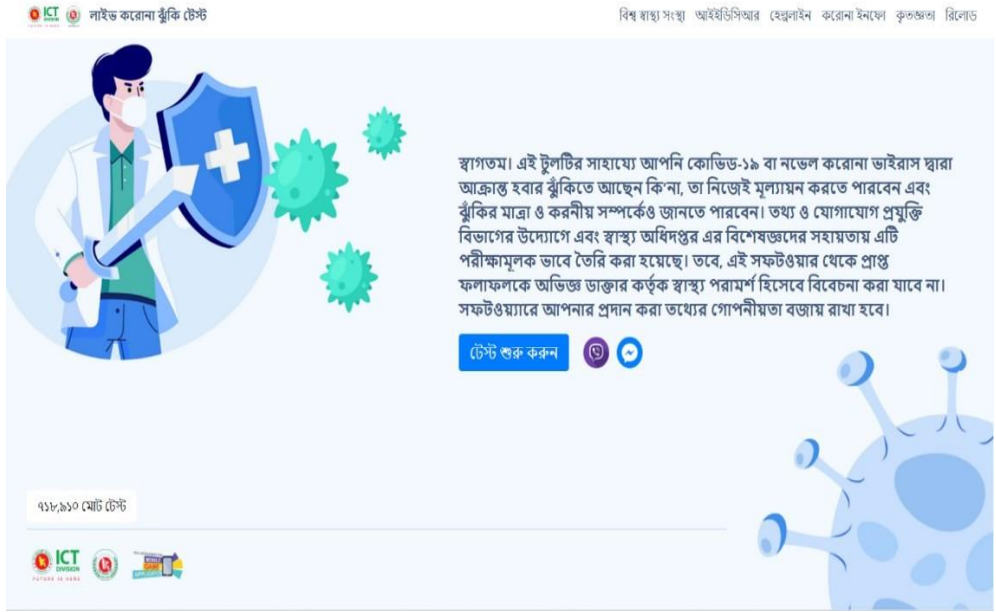

Figure 4. User Interface of Surokkha application.

A data dashboard was developed, an information management tool that visually tracks, analyses, and displays key indicators regarding the status of COVID-19 in Bangladesh. Different types of metrics and key data points were generated and will be displayed in the dashboard to have detailed and bird's eye views. Location and mapping features were the essential building blocks for this platform. Different types of metrics and key data points were generated and will be displayed in the dashboard to have detailed and bird's eye views. We have planned to make this system with a mapping first approach. We have chosen the Google maps feature to build an efficient and helpful system.

### 3.4. Hello Doctor

Hello Doctor is the most innovative mobile application to get medical consultancy. It helps by Telemedicine to make a video call to doctor to get medical consultancy [10]. Appointments Get an appointment with experienced Doctors anywhere, 24/7. E-prescription Get your prescription in your app and through SMS. Medicine Reminder Track your medicines and get reminders every time.

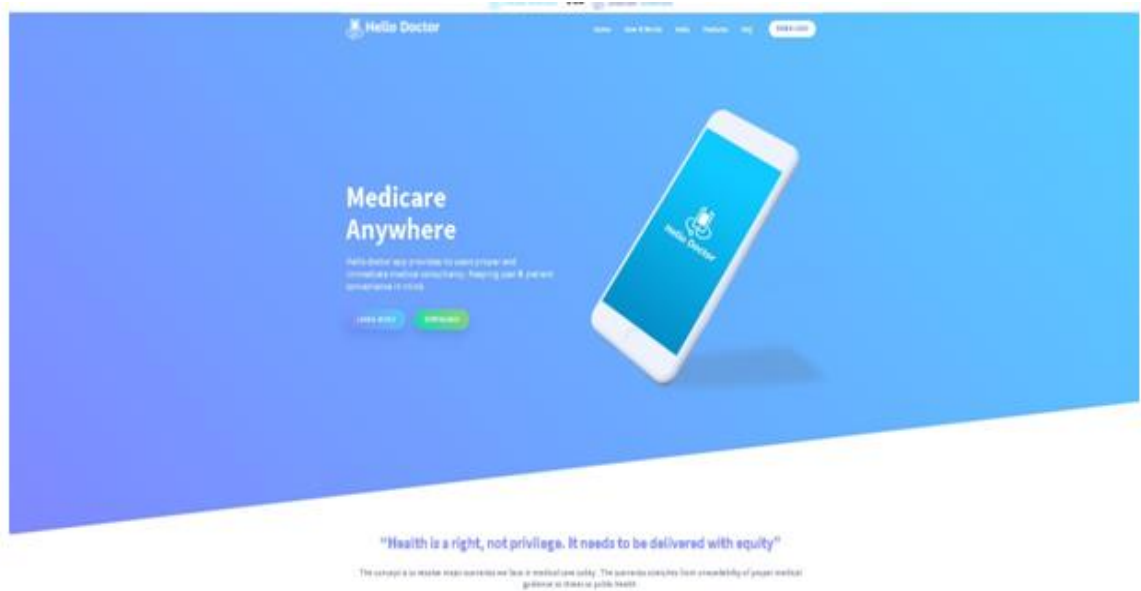

Figure 5. Hello doctor web dashboard

Utilities Health tips, Emergency medical actions, Medicines for primary pre-cautions. Clean design and attractive design will create a great user experience. User-friendly experience is the best user-friendly Medicare application.

### 3.5. Corona Tracer

Corona Tracer-BD is an initiative by the ICT Division and DGHS to bring the people of Bangladesh together in the combined fight against COVID-19 [11]. Shohoz Ltd. powers the technology. This app utilizes a Bluetooth signal to understand if you are near another Tracer app user. It will help you identify whether you are at risk of COVID-19 infection by checking if you have been in recent contact with an infected individual. If your case seems risky, you will be able to seek medical help at the earliest and go into self-isolation.

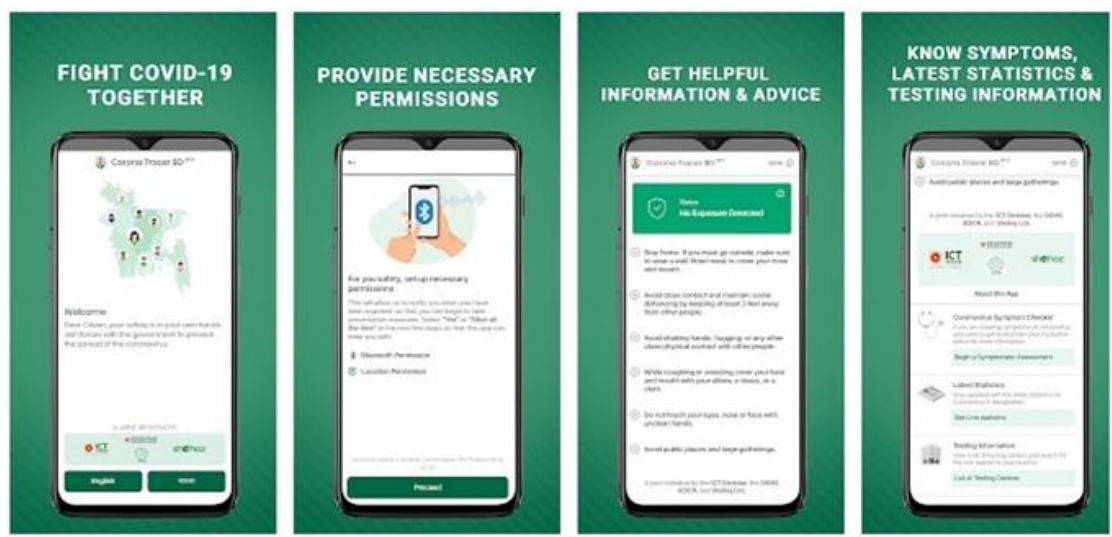

Figure 6. Corona Tracer BD App on Play store.

It provides information on various topics such as symptoms, what to do in case of suspected infection, and locating nearby health and testing units. In case of a suspected infection, the citizen can check if the symptoms are compatible with COVID-19's, and if so, they will be instructed and sent to the nearest primary health unit. The Government's official news area for viewing the latest statistics on the pandemic.

### 3.6. Education for the Nation

The Innovation Design and Entrepreneurship Academy (IDEA) project of Bangladesh Computer Council (BCC) under the ICT Division will provide technical support to conduct the online classes [12].

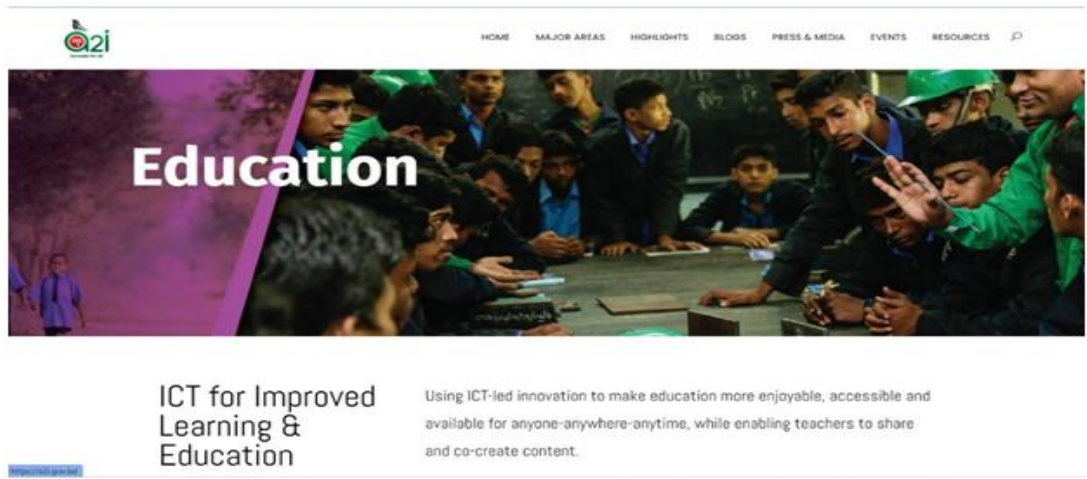

Figure 7. Education for nation website dashboard

Classes of ninth, tenth, eleventh, and twelfth grades will be taken online. Each class will be continued for 60 minutes with 45 minutes of teaching and the remaining 15 minutes of question and answer (Q/A) sessions. Students will be provided with the routine of the classes and the necessary materials in advance. The Start-up Bangladesh-IDEA project will also create an online platform to deliver the same service through online management and with the help of start-ups in different educational institutions across the country centring on the education sect.

### 3.7. BSMMU-a2i Specialized Telehealth Centre

BSMMU-a2i Specialized Telehealth Centre: During the Covid-19 crisis, citizens were being deprived of regular healthcare services [13]. In this regard, the Bangabandhu Sheikh Mujib Medical University (BSMMU), in collaboration with ICT Division, launched the BSMMU-a2i Specialized Telehealth Centre (09611677777) to provide medical advice to citizens of the country through video and audio calls. By this platform, all citizens of Bangladesh may get medical suggestions and information.

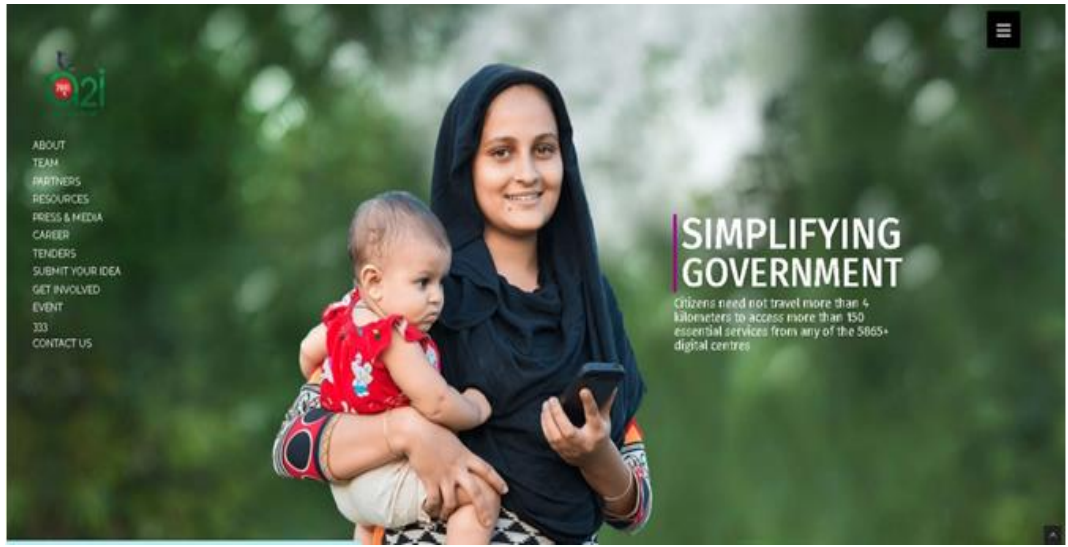

Figure 8. BSMMU-a2i specialized telehealth centre website dashboard

### 3.8. The Crowdfunding Platform Ek Desh

The country's first Crowdfunding platform Ek Desh has been launched to facilitate Zakat and donations distribution for people, quickly from anywhere, anytime through banking channels [14]. The technology-based Ek Desh platform has been launched to ensure financial assistance and social security to the underprivileged people and small businesses affected by the Corona crisis. Through the Ek Desh platform, people from different walks of life across the country can pay Zakat or small financial donations to selected Government and non-government organizations, including the Prime Minister's Relief Fund and the Islamic Foundation.

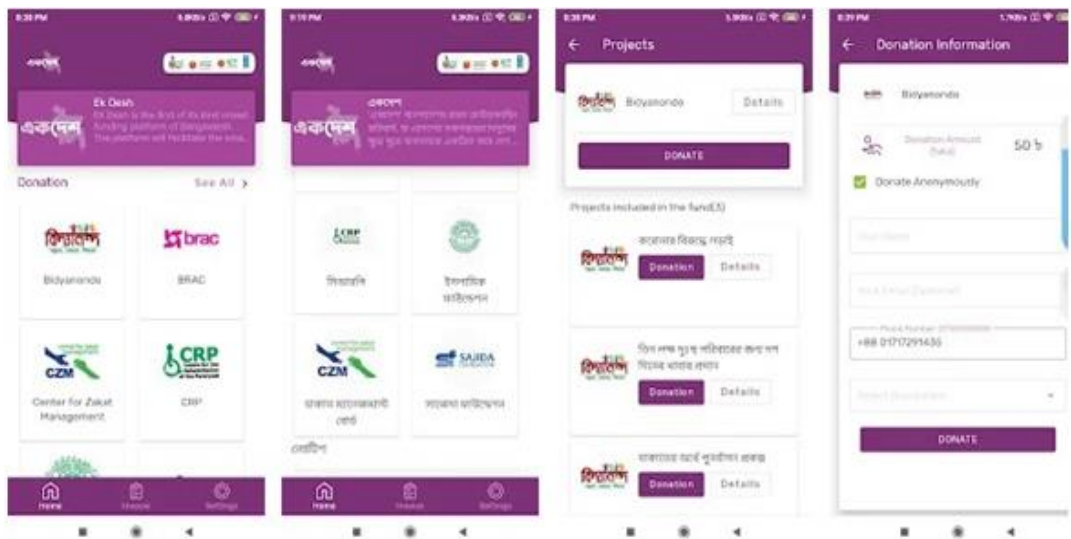

Figure 9. Ek Desh app user interface available on the Play store

### 3.9. Food for Nation and Phone-E-Nittoponno

A public-private coordination group has been set up under the initiative of a2i's e-shop, and a logistic network has been established coordinating 116 e-commerce and logistics companies, e-CAB, and Dhaka Divisional Administration [15]. 1000 pharmacies and more than 1 lakh registered points/shops, including grocery stores, have been prepared. Marketing of agricultural products of local farmers has also been initiated across the country, and the Food for Nation platform's activities are underway in coordination with the Ministry of Agriculture.

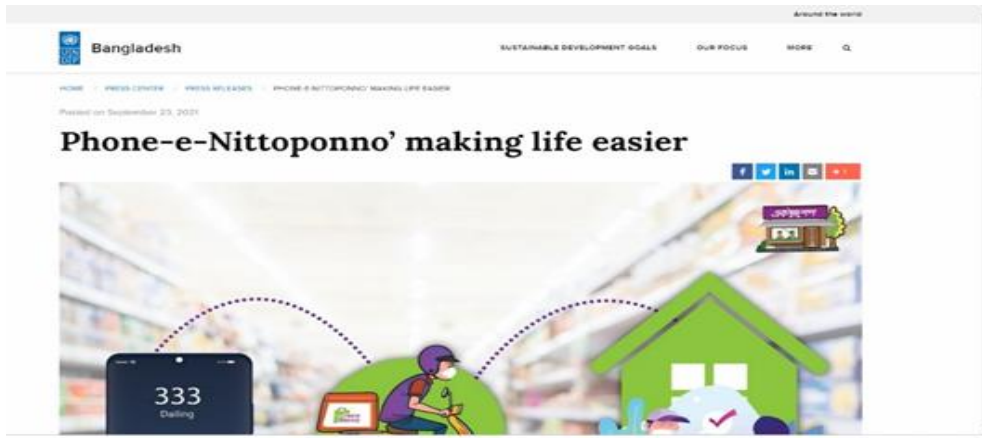

Figure 10. Food for nation related news report on newspaper

## 4. DISCUSSION

In this section, we describe a comparative analysis of the approaches adopted by Bangladesh in comparison with the rest of the world. Like other countries, Bangladesh has taken various steps to prevent coronavirus. The most challenging task in Bangladesh was to bring people from real life to virtual life and get services from there. Now we believe that digital technologies join a long line of public-health innovations that have been at the heart of disease-prevention-and-containment strategies for centuries. The unprecedented human and economic needs presented by COVID-19 are driving scale and speed development and adoption of new digital technologies. We have highlighted the potential of digital technology capable of tracing and identifying cases in rapid identification and monitoring travel patterns lockdown time and dimensions of public-health messaging to support epidemiological intelligence with online datasets. All the digital platforms used in this pandemic situation are given many benefits all over Bangladesh. There has a percentage of these applications in which information are available online source is below.

Table 1. The percentage of users befitted from digital technology [16]

| Name | Benefit Percentage |
|---|---|
| Boithok | 66% |
| Surokkha | 86% |
| Live corona test | 84% |
| Corona Action Bot | 84% |
| Hello Doctor | 84% |
| Covid 19 Tracker | 80% |
| Ek Desh | 82% |
| Corona Tracer | 72% |

Table 1 shows the percentage of Boithok at 66%, Surokkha at 86%, Live Corona test 84%, Corona Action bot 84%, Hello Doctor 84%, COVID-19 Tracer 80%, Ek Desh 82%, and Corona Tracer 72% at the point of user benefits. The highest percentage at the Surokha application is about to 86%, and the lowest at the Boithok application is about 66%. A graphical representation of these percentage with services in Figure 11 shows the user benefits and another line for services. This graph shows us a ratio between that's above all the application.

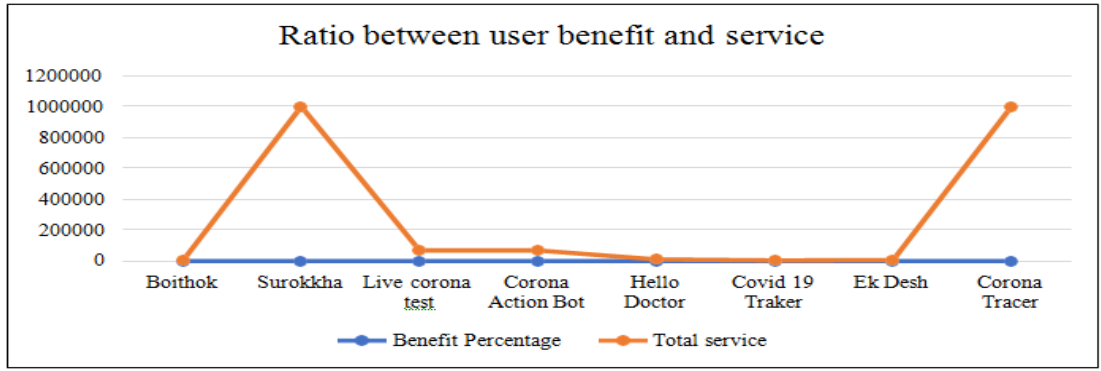

Figure 11. The ratio between user benefit and services [16]

The widespread use of digital solutions remains a barrier. Bangladesh has increased the number of users of this COVID worldwide application internet, mobile, and different types and websites. Now for universal improvement and to lead a safer life in Bangladesh, mobile applications and websites are taking a virtual place in various innovations. These COVID worldwide following mobile applications and websites are digital innovations. Public health may become increasingly digital in the future, and this case and worldwide preparation plans, recognizing the importance of digital technology, have become urgent. Technology companies such as key stakeholders in the digital field, rather than partners, should be long-term partners preparing only when the emergency is underway. Viruses know no boundaries and, increasingly, neither digital technology nor information. Coordination of international strategies for controlling, evaluating, and using digital technology is needed to strengthen epidemic management and future preparedness for COVID-19 and other infectious.

## 5. CONCLUSION

In this research, all the mobile applications and websites are evaluated through Digital Technology in the public health response to the COVID-19 the Bangladesh perspective. All those applications and websites are on acceptation, but all applications have to improve some informatics issues. In this study, we have systematically reviewed the digital technology initiatives that have been taken across the world and Bangladesh for the appeasement of the pandemic expansion of COVID-19. Based on our relative exploration, we propose some areas where Bangladesh can reduce the pandemic spread of COVID-19. After analysis, we think that some of the digital technology that has been taken and confirmed helpful in the global aspect may be possible in Bangladesh. To acquire this, we think a collaboration between Government and private sectors will be needed. Our research study has several limitations, but at the same time, provides some capital for future research. First, we used specific keywords to search for the relevant material applications and websites. Second, it may be that the elements that we have used may have some biases in their reporting data. Third, we think timely and up-to-date data are the key to most digital technology that we have recognized our study for in this paper. Therefore, upcoming works would require collecting exact data applying multiple processes and not just depending on testing facilities. This research validates those applications created to better human life in the COVID-19 pandemic and still gives us helpful information-friendly services with their ability.

## AUTHORS

**Samrat Kumar Dey** is a Lecturer at the School of Science and Technology (SST), Bangladesh Open University (BOU), Bangladesh. He has received his M.Sc. and B.Sc. in Computer Science and Engineering from Military Institute of Science and Technology (MIST), BUP, Bangladesh, and Patuakhali Science and Technology University (PSTU), Bangladesh in 2019 and 2015 respectively. His research interest includes Visual Data Analytics, Health Informatics, Information visualization, HCI, Machine Learning, and Deep Learning. Mr. Samrat has published more than thirty (35+) research articles in reputed journals and conferences. Also, he has served as a technical committee member and reviewer for different international journals and conferences worldwide.

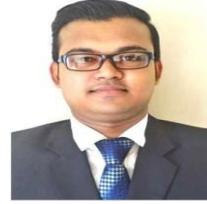

**Khaleda Mehrin** is an Assistant Programmer, UITRCE, BANBEIS (Bangladesh Bureau of Educational Information and Statistics), Ministry of Education, Bangladesh. She has received M.Sc.in Computer Science from Jahangirnagar University (JU), Bangladesh, and B.Sc. from Patuakhali Science and Technology University (PSTU), Bangladesh in 2016 and 2012 respectively. As an expert IT officer, she is contributing to the field of Information and Technology in Bangladesh through her job.

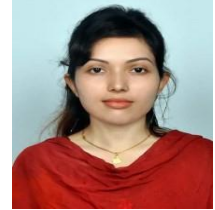

**Lubana Akter** is an Undergrad student of Computer Science and Engineering at the Dhaka International University, Bangladesh. She is the Regional Winner at the NASA Space Apps Challenge 2021. Her research interest mainly focuses on Data analysis, Machine learning, and big data. She is Skilled in software development, python, and Object-oriented programming (OOP).

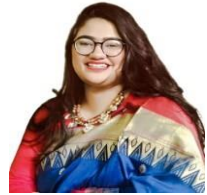

**Mshura Akter** is a student of Dhaka International University at the Department of Computer Science and Engineering. She is mainly interested in the Software Complexity, Network Security, and IoT research domain.

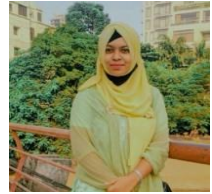